\documentclass[aps,prx,twocolumn,amsmath,amssymb,superscriptaddress,floatfix]{revtex4-2}

\usepackage{mathtools}
\usepackage{multirow}
\usepackage{xcolor}
\usepackage{bm}
\usepackage{amsfonts,amssymb,amsmath}
\usepackage{graphicx,dcolumn,bm,xcolor,braket,slashed}
\usepackage{times} 
\usepackage{comment}
\usepackage{array}
\usepackage{textcomp}
\usepackage[normalem]{ulem}
\usepackage{dsfont}
\usepackage{kotex}

\usepackage[title,toc,titletoc,page]{appendix}

\newcommand{\be}{\begin{eqnarray}}
\newcommand{\ee}{\end{eqnarray}}
\newcommand{\bbm}{\begin{bmatrix}}
\newcommand{\ebm}{\end{bmatrix}}
\newcommand{\bpm}{\begin{pmatrix}}
\newcommand{\epm}{\end{pmatrix}}

\begin{document}
\title{Quantum geometry and RKKY in flat bands}

\author{Chang-geun Oh}
\email{cg.oh.0404@gmail.com}
\affiliation{Department of Applied Physics, The University of Tokyo, Tokyo 113-8656, Japan}
\author{Makoto Shimizu}
\affiliation{Graduate School of Engineering Science, The University of Osaka, Toyonaka, Osaka 560-8531, Japan}
\affiliation{Department of Physics, Graduate School of Science, Kyoto University, Kyoto 606-8502, Japan}
\author{Youichi Yanase}
\affiliation{Department of Physics, Graduate School of Science, Kyoto University, Kyoto 606-8502, Japan}
\author{Shuichi Murakami}
\affiliation{Department of Applied Physics, The University of Tokyo, Tokyo 113-8656, Japan}
\affiliation{RIKEN Center for Emergent Matter Science (CEMS), 2-1 Hirosawa, Wako, Saitama, 351-0198, Japan}
\affiliation{
International Institute for Sustainability with Knotted Chiral Meta Matter (WPI-SKCM$^2$),Hiroshima University, 1-3-1 Kagamiyama, Higashi-Hiroshima, Hiroshima 739-8526, Japan}

             
\begin{abstract}

Flat conduction bands quench the group velocity and thus challenge conventional, dispersion-driven pictures of the Ruderman–Kittel–Kasuya–Yosida (RKKY) interaction, where localized moments are coupled via an effective exchange mediated by conduction electrons. 
Here we show that RKKY interactions in the flat-band limit are not extinguished by the vanishing group velocity but are instead mediated by the quantum geometry of Bloch states.
Starting from a microscopic RKKY derivation, we demonstrate that the Brillouin-zone–averaged quantum metric controls the long-wavelength structure of the static susceptibility, thereby determining the magnetic correlation length and the spin stiffness.
As a result, the finite spatial spread of Wannier functions provides an effective long-range coupling channel even when single-particle dispersion is absent.
Furthermore, we establish the general principle that the ordering temperature is governed by the quantum metric  in finite and low-dimensional samples, effectively circumventing the thermodynamic-limit constraint of the Mermin–Wagner theorem. Specifically, our theoretical investigation reveals that increasing the quantum metric enhances magnetic rigidity and leads to a corresponding rise in the critical temperature within finite-sized systems.
\end{abstract}
\maketitle

\textit{Introduction.}
The recognition that quantum geometry dictates electronic properties beyond the paradigm of band structure has emerged as a central theme in modern condensed matter physics~\cite{gao2025quantum, yu2025quantum,torma2023essay}. The quantum geometric tensor (QGT), along with its symmetric and antisymmetric components—the quantum metric and the Berry curvature~\cite{provost1980riemannian, berry1989quantum}—has been shown to be directly linked to a wide range of physical observables.
While the antisymmetric part is fundamental to topological physics~\cite{nagaosa2010anomalous, Xiao_2010_RMP}, the symmetric part has recently been found to govern phenomena such as superconducting properties~\cite{peotta2015superfluidity,torma2022superconductivity,oh2025role,chen2024ginzburg}, optical responses~\cite{oh2025universal,oh2025color,cook2017design, de2017quantized,ahn2022riemannian, morimoto2016topological}, Landau‐level spreading~\cite{rhim2020quantum,hwang2021geometric,oh2024bilayer}, electron scattering with disorders~\cite{oh2024thermoelectric} and phonons~\cite{yu2024non}, Klein tunneling~\cite{han2026klein}, exciton condensate~\cite{verma2024geometric}, bulk-interface correspondence in singular flat bands~\cite{oh2022bulk,kim2023general}, and magnetic properties~\cite{Wu2020,Kitamura2024,oh2025magnetic,shimizu2026magnetic}.

Traditionally, the Ruderman–Kittel–Kasuya–Yosida (RKKY) interaction~\cite{ruderman1954indirect,kasuya1956theory,yosida1957magnetic}, the mechanism by which localized magnetic moments couple indirectly through conduction electrons, is understood to be governed by electronic dispersion and the structure of the Fermi surface. 
In this classical picture, the group velocity of electrons form Friedel oscillations~\cite{friedel1958metallic}, resulting in a momentum-dependent static susceptibility $\chi^0(\bm{q})$ that generates effective exchange interactions. 
However, when the conduction band is perfectly flat, this dispersion-based mechanism collapses. From a naive perspective, the electrons are considered strictly localized, rendering the susceptibility $\bm{q}$-independent and thus unable to mediate spatially structured exchange between localized moments.

This work demonstrates that such intuition is incomplete. We show that the RKKY interaction does not vanish in the flat-band limit; rather, its origin lies in the quantum geometry of the band. 
We reveal that the RKKY susceptibility is acutely sensitive to the quantum-mechanical overlap of Bloch wavefunctions, from which we identify that the Brillouin-zone-averaged quantum metric governs the long-wavelength structure of the susceptibility, the magnetic correlation length, and the spin stiffness. 
Physically, even in the absence of dispersion, the finite spatial spread of the Wannier functions ensures that electrons provide an effective long-range mediation channel between localized spins without requiring net electronic motion.

Considering the experimental context of low-dimensional systems, we establish a general principle that the critical temperature $T_c$ in finite-sized samples is determined by the quantum geometry. The Mermin–Wagner theorem forbids spontaneous breaking of continuous symmetries at finite temperature in the thermodynamic limit because long-wavelength thermal fluctuations destroy long-range order in one- and two-dimensional systems~\cite{mermin1966absence,hohenberg1967existence}. 
Nevertheless, we show that for finite systems the spin stiffness determined by the quantum metric provides sufficient rigidity to suppress these fluctuations, yielding an observable $T_c$ that scales with the geometric properties.

\textit{Formalism.}
We consider a tight-binding model of conduction electrons coupled to localized magnetic moments on a $d$-dimensional lattice, described by the Hamiltonian:
\begin{align}
    &H=H_0+H_K, \nonumber\\
    &H_0=\sum_{i,j,\sigma}t_{ij}c^\dagger_{i\sigma}c_{j\sigma}=\sum_{\bm k , \alpha,\beta,\sigma}c^\dagger_{\alpha\sigma}(\bm{k})H^{\alpha\beta}(\bm{k})c_{\beta\sigma}(\bm{k}), \nonumber  \\
    &H_K=J_K\sum_{i}\bm{S}_i\cdot\bm{s}_i, \quad \bm{s}_i \equiv \frac{1}{2}c^\dagger_{i}\bm{\sigma} c_i, 
\end{align}
where $\bm{\sigma} = (\sigma^x, \sigma^y, \sigma^z)$ is the vector of Pauli matrices, $\bm{S}_i$ represents the localized magnetic moment at site $i$, $J_K$ describes the Kondo exchange coupling, and $c_{i} = (c_{i\uparrow}, c_{i\downarrow})^T$ is the spinor for the conduction electrons. The operators $c^\dagger_{i\sigma}$ ($c_{i\sigma}$) create (annihilate) an electron at site $i$ with spin $\sigma$. In momentum space, $c_{\alpha\sigma}(\bm{k})$ denotes the annihilation operator with wavevector $\bm{k}$, spin $\sigma$, and internal degrees of freedom $\alpha$ (e.g., sublattice indices).

The effective action for localized magnetic moments to second order in $J_K$, obtained by integrating out the conduction electron degrees of freedom, leads to the following RKKY Hamiltonian~\cite{SM}
\begin{align}
    H_{\mathrm{eff}}=-\frac{J_K^2}{4} \sum_{i,j}\chi^0_{ij}\bm{S}_i\cdot\bm{S}_j.
\end{align}
where $\chi^0_{ij}$ is the bare (non-interacting) susceptibility of the conduction electrons. 
For a translationally invariant lattice, $\chi^0_{ij}$ depends only on the relative coordinate $\bm{r}_i - \bm{r}_j$ and is related to its momentum-space counterpart $\chi^0(\bm{q})$ via the Fourier transform as 
\begin{align}
    \chi^0_{ij} = \int \frac{d\bm{q}}{(2\pi)^d} e^{i\bm{q} \cdot (\bm{r}_i - \bm{r}_j)} \chi^0(\bm{q}).
\end{align}
The momentum-dependent susceptibility $\chi^0(\bm{q})$ is calculated using the Bloch eigenstates of the non-interacting Hamiltonian $H_0$. It is given by~\cite{oh2025magnetic,Kitamura2024,SM} 
\begin{align}
\chi^0(\bm{q})=\sum_{n,m}\int \frac{d\bm{k}}{(2\pi)^d}
F_{nm}(\bm{k},\bm{q})
\bigl|\langle u_n(\bm{k+q})|u_m(\bm{k})\rangle\bigr|^2 \label{eq:chi_bare_tot}
\end{align}
where $f(\epsilon) = (\exp[(\epsilon - \mu) / T] + 1)^{-1}$ with $\mu$ as the chemical potential, $F_{nm}(\bm{k,q})=[f(\epsilon_m(\bm{k}))-f(\epsilon_n(\bm{k+q}))]/[\epsilon_n(\bm{k+q})-\epsilon_m(\bm{k})]$, and $\epsilon_n(\bm k)$ and $\lvert u_n(\bm k)\rangle$ are the energy and the Bloch eigenstate defined by $H_0\lvert u_n(\bm k)\rangle=\epsilon_n(\bm k)\lvert u_n(\bm k)\rangle$.

The form factor $\bigl| \langle u_n(\bm{k}+\bm{q}) | u_m(\bm{k}) \rangle \bigr|^2$ represents the quantum-mechanical overlap between Bloch states, which encodes the quantum geometry of the bands. As highlighted in recent studies~\cite{oh2025magnetic,Kitamura2024}, this geometric contribution implies that magnetic order can be fundamentally driven by the quantum geometry of the Bloch wavefunctions, rather than being solely a consequence of the band dispersion.

\begin{figure}[t]
\includegraphics[width=80mm]{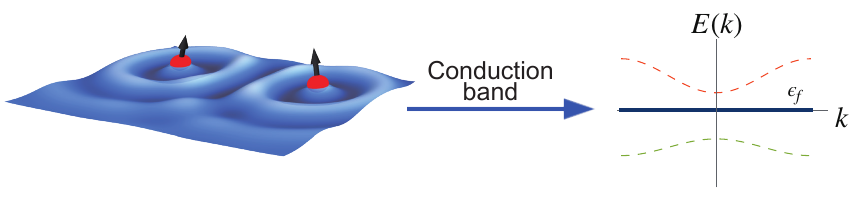} 
\caption{
\textbf{Schematic illustration of localized spins (red) interacting through a flat conduction band.} The right panel depicts the band structure $E(k)$ featuring a flat band (solid blue) at the Fermi level, separated from remote bands (dashed lines) by an energy gap $\Delta$. 
\label{fig1}}
\end{figure}

\textit{Flat band limit.}
We consider the scenario where the Fermi level $\mu$ is tuned exactly to the energy $\epsilon_f$ of the flat band, as illustrated in the schematic of Fig. 1.
In the low-temperature limit ($T \ll \Delta$, where $\Delta$ is the energy gap to the nearest band), the susceptibility $\chi^0(\bm q)$ is dominated by intra-band transitions within the flat band. Assuming $\epsilon_f = \mu$, the Lindhard factor simplifies to $F_{ff} = 1/(4T)$, and the susceptibility is given by
\begin{align}
\chi^0(\bm q) = \frac{1}{4T} \int \frac{d\bm k}{(2\pi)^d} \bigl| \langle u_f(\bm{k+q}) | u_f(\bm{k}) \rangle \bigr|^2.
\end{align}
This expression reveals that the behavior of $\chi^0(\bm{q})$ is fundamentally dictated by the quantum geometry of the flat band.

In the geometrically trivial limit, where the flat band originates from a single orbital per unit cell without internal degrees of freedom, rendering the Bloch states $\bm{k}$-independent ($|u_f(\bm{k}) \rangle = |u_f \rangle$). In this scenario, the overlap $|\langle u_f(\bm{k+q}) | u_f(\bm{k}) \rangle|^2$ is unity for all $\bm{q}$, making $\chi^0(\bm{q})$ a constant and $\chi^-_{ij}$ to vanish for $i\neq j$. Physically, this corresponds to the atomic limit of zero hopping ($t_{ij}=0$), where conduction electrons are sharply localized at individual sites. Being immobile and lacking inter-site connectivity, these electrons are unable to mediate a spatially dependent exchange interaction between localized magnetic moments.

Conversely, in systems with multiple sublattices or orbitals, a flat band can emerge from destructive interference despite the presence of finite hopping~\cite{leykam2018artificial,mielke1991ferromagnetism,tasaki1992ferromagnetism}. Although the group velocity vanishes ($\bm{v}_{\bm{k}} = \nabla_{\bm{k}} \epsilon_{\bm{k}} = 0$), the Bloch wavefunctions possess a non-trivial momentum-space texture. Consequently, the overlap $|\langle u_f(\bm{k+q}) | u_f(\bm{k}) \rangle|^2$ becomes $\bm{q}$-dependent and is strictly less than unity for $\bm{q} \neq 0$. This results in a well-defined maximum at $\bm{q} = 0$, which is essential for stabilizing ferromagnetic order.

Physically, the non-trivial geometry implies that the corresponding Wannier functions are not delta-like but exhibit a finite spatial spread across multiple sites~\cite{marzari1997maximally,resta2011insulating}. It is precisely this Wannier function overlap that allows the immobile conduction electrons to bridge the distance between localized moments, mediating an effective RKKY interaction even in the absence of band dispersion.

To further quantify the geometric contribution to the magnetic interaction, we examine the behavior of the susceptibility in the long-wavelength limit ($q \to 0$). By Taylor expanding the Bloch state $| u_f(\bm{k}+\bm{q}) \rangle$ around $\bm{q}=0$, the susceptibility is given by
\begin{align}
\chi^0(\bm{q}) \approx \chi^0(\bm{0}) - \frac{1}{4T} \sum_{a,b} q^a q^b \bar{g}_{ab}, \label{eq:suscept}
\end{align}
where $\chi^0(\bm 0 )=1/(4T)$, and $\bar{g}_{ab} = \int \frac{d^d\bm{k}}{(2\pi)^d} g_{ab}(\bm{k})$ is the Brillouin-zone-averaged quantum metric of the flat band. The local quantum metric $g_{ab}(\bm{k})$ is defined as $g_{ab}(\bm{k}) = \mathrm{Re} \bigl[ \langle \partial_a u_f(\bm{k}) | \partial_b u_f(\bm{k}) \rangle - \langle \partial_a u_f(\bm{k}) | u_f(\bm{k}) \rangle \langle u_f(\bm{k}) | \partial_b u_f(\bm{k}) \rangle \bigr]$,
where $\partial_a \equiv \partial/\partial k_a$.

Assuming that the peak of $\chi^0(\bm{q})$ at $\bm{q}=0$ leads to a ferromagnetic ground state for the localized moments, Eq.~(\ref{eq:suscept}) identifies the average quantum metric as a square of the magnetic correlation length as $\xi_{ab}^2 \equiv \bar{g}_{ab}$. 
Furthermore, within linear spin-wave theory, the magnon dispersion is directly governed by the quantum geometry, as derived in the Supplemental Material~\cite{SM}:
\begin{align}\hbar \omega(\bm{q}) = \sum_{a,b}D_{ab} q_aq_b, \quad D_{ab} = S \frac{J_K^2 \chi^0(\bm 0)}{2} \bar{g}_{ab},\end{align}
where $D$ is the spin stiffness, which characterizes the energy cost of long-wavelength magnetic fluctuations.
Thus, the stability of the magnetic order emerges as a purely geometric effect: a larger quantum metric implies a more significant Wannier function overlap, which enhances the spin stiffness and reinforces the ferromagnetic phase against fluctuations.
Note that recent studies on magnetism in itinerant electronic systems have also demonstrated that the spin stiffness is governed by the quantum geometry of electronic wave functions~\cite{kang2024quantum}.

\textit{Dimensionality and geometric stabilization.}
The stability of magnetic order is profoundly influenced by the dimensionality $d$ of the system. 
In $d>2$, the contribution of long-wavelength magnons to the magnetization is infrared convergent, allowing finite-temperature magnetic order. Mean-field theory (MFT) then provides a useful baseline estimate of the ordering temperature $T_c^{\mathrm{MF}}$, although critical fluctuations can still modify its quantitative value in finite dimensions~\cite{goldenfeld2018lectures}.
Within MFT, the ordering temperature is determined by
\begin{align}
1=
\frac{S(S+1)J_K^2}{6T_c^{\mathrm{MF}}}
\chi^0(\bm q_\star,T_c^{\mathrm{MF}}),
\end{align}
where $\bm q_\star$ denotes the wave vector at which $\chi^0(\bm q)$ is maximal. For a conventional dispersive metal, where $\chi^0(\bm q_\star,T)$ is approximately temperature independent near the transition, this condition gives $T_c^{\mathrm{MF}}\propto J_K^2$.
By contrast, in the ideal isolated flat-band limit, the ferromagnetic instability occurs at $\bm q_\star=\bm 0$ and $\chi^0(\bm 0,T)=1/(4T)$. The mean-field condition then yields $T_c^{\mathrm{MF}} \propto J_K$.
At this mean-field level, the transition scale is independent of the averaged quantum metric because $\bar g_{ab}$ controls the momentum dependence of $\chi^0(\bm q)$ near $\bm q=\bm 0$, rather than its peak value $\chi^0(\bm 0)$.

In contrast, in $d\leq 2$, MFT provides only an upper estimate of the ordering scale~\cite{goldenfeld2018lectures}.
Furthermore, the Mermin-Wagner theorem states that for infinite systems with continuous symmetry, long-range magnetic order is strictly forbidden at any finite temperature because long-wavelength magnons lead to an infrared divergence that destroys the spontaneous magnetization~\cite{mermin1966absence,hohenberg1967existence}. 
However, in practical scenarios involving finite-sized systems of length or width $L$, a natural infrared cutoff $q_{\min} \sim 2\pi/L$ is introduced. This cutoff prevents the divergence of magnons, allowing for the emergence of ferromagnetic state at finite temperatures~\cite{palle2021physical,jenkins2022breaking,tomita2014finite}. In this finite-size regime, the stability is no longer determined by the zero-momentum susceptibility alone, but by the temperature-dependent spin stiffness $D(T) = \frac{S J_K^2 \bar{g}}{8k_B T}$, which is intrinsically tied to the quantum geometry.

For a one-dimensional system, solving the magnetization depletion condition $\delta m(T_c) \sim S$ yields (see Supplemental Material~\cite{SM} for the derivation.)
\begin{align}
T_c^{(1D)} \simeq \frac{\pi J_K S \sqrt{\bar{g}}}{2\sqrt{L}a},
\end{align}
where $L=Na$ denote the size of the system. Here, $N$ is the number of unit cells and $a$ is the lattice constant. 
The transition temperature scales as $L^{-1/2}$, recovering the Mermin-Wagner result ($T_c \to 0$) in the thermodynamic limit. Crucially, $T_c$ is directly proportional to the square root of the average quantum metric $\sqrt{\bar{g}}$, demonstrating that the geometric spread of the Wannier functions is the key factor providing stability against 1D thermal fluctuations.

Similarly, in two dimensions, $T_c$ is given by
\begin{align}
T_c^{(2D)} \simeq \frac{\sqrt{\pi} J_K S ( \det \bar{g} )^{1/4}}{ 2a\sqrt{\ln [L/(2a)]}}.
\end{align}
While $T_c$ still vanishes in the thermodynamic limit $L\to\infty$, its logarithmically slow decay implies a sizable finite-size ordering scale and robust magnetic correlations even in large systems.
These results underscore that the stability of magnetism is governed by the quantum geometry.

\textit{1D toy model.}
\begin{figure}[t]
\includegraphics[width=80mm]{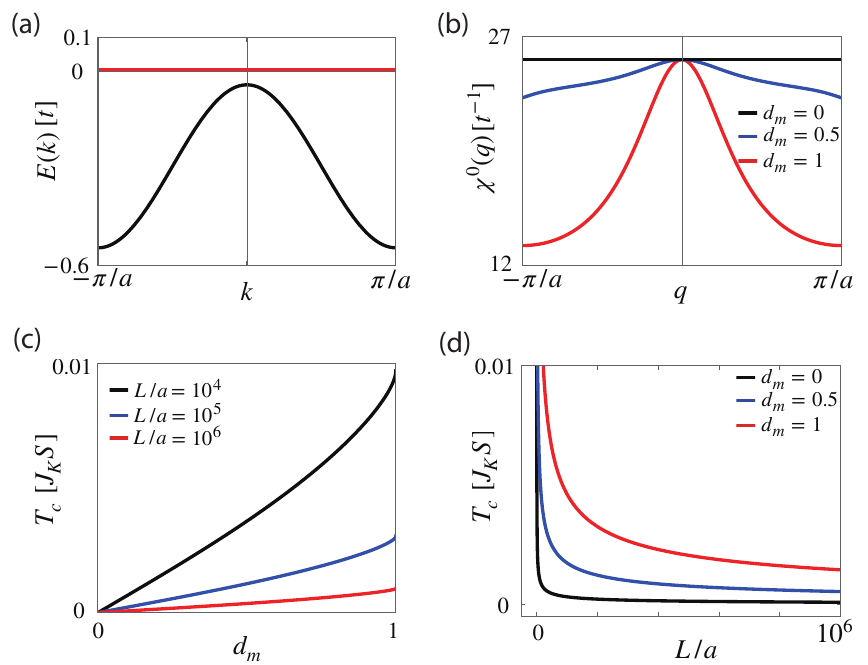} 
\caption{
\textbf{Geometric stabilization of ferromagnetic order in a 1D RKKY model.} (a) Energy dispersion of $H_c^{1D}$. The energy spectrum remains invariant under the tuning of the geometric parameter $d_m$. (b) Static susceptibility $\chi^0(q)$ for several values of $d_m=0$, 0.5 and 1. (c) Critical temperature $T_c$ as a function of $d_m$ for fixed system sizes ($L = 10^4, 10^5, 10^6$). (d) $T_c$ as a function of $L$ for $d_m = 0.1$, 0.5 and 1. Here, we fixed $\Delta=0.3$ for (b-c) and $T=0.01$ for (b).
\label{fig2}}
\end{figure}

To verify the role of quantum geometry in stabilizing the ferromagnetic order of localized electrons, we investigate the following Hamiltonian for the conduction electrons in momentum space: $H_c^{1D}(k) = t\bigg(\frac{\cos (ka) - 1 - 2\Delta^2}{8} \tau_0 + \frac{d_m\Delta^2\sqrt{1-d_m^2}}{2}\tau_x  + \frac{d_m\Delta\sin(ka/2)}{2}\tau_y + \frac{1 + 2\Delta^2 - 4d_m^2\Delta^2 - \cos (ka)}{8}\tau_z\bigg)$, where $\tau_0$ and $\tau_i$ ($i=x,y,z$) denote the identity and Pauli matrices, respectively.
Here, $t>0$ sets the overall energy scale of the model and may be regarded as the characteristic hopping amplitude. The real, dimensionless parameters $\Delta$ and $d_m$ characterize the band separation and quantum geometry, respectively.
This Hamiltonian yields a perfectly flat band at $E_1(k) = 0$ and a dispersive band $E_2(k) = \frac{t}{4} (-1 - 2 \Delta^2 + \cos (ka))$, as shown in Fig.~2(a). We set the chemical potential to the flat-band energy, $\mu=0$, such that the flat band lies at the Fermi level.

Notably, the energy eigenvalues of $H_c(k)$ are completely independent of $d_m\in [0, 1]$. 
The parameter $d_m$ controls the momentum-space texture of the Bloch wave functions, thus it tunes the quantum metric, while preserving the band dispersion.
It should be noted that the quantum metric also depends on the orbital positions~\cite{oh2026orbital}. For simplicity, we assume fixed orbital positions where the two sublattices are separated by a distance of $a/2$.
Consequently, for a fixed $\Delta$, the stability of the ferromagnetic phase is governed exclusively by geometric tuning via $d_m$. 
As $d_m$ increases with a fixed $\Delta$, the averaged quantum metric $\bar{g}$ grows. 
The increased $\bar{g}$ leads to a sharper peak of the susceptibility $\chi^0(\bm{q})$, and results in a higher critical temperature $T_c$ for a fixed system size $L$, as illustrated in Figs.~2(b) and 2(c).
The system size $L$ dependence of $T_c$ for $d_m=0.1, 0.5, 1$ is further detailed in Fig.~2(d). 
These results confirm that in the flat conduction band limit, the geometric structure of conduction electronic states governs the RKKY interaction and provides the necessary stiffness to stabilize the magnetic order against thermal fluctuations.

\textit{2D Lieb lattice model.}
\begin{figure}[t]
\includegraphics[width=80mm]{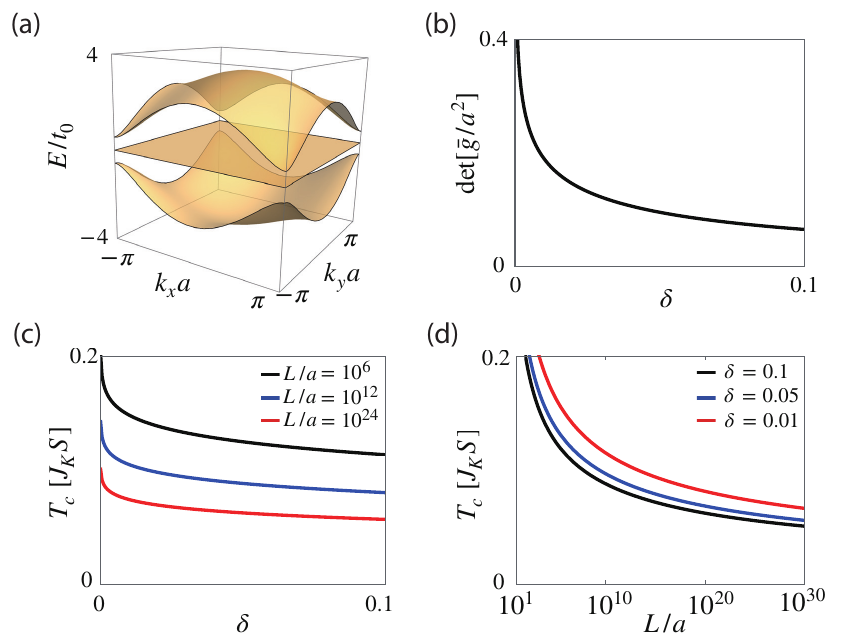} 
\caption{
\textbf{Geometric stabilization of ferromagnetic order in a 2D RKKY model.} (a) Energy dispersion of $H_c^{2D}$ with $\delta=0.4$. (b) Determinant of the averaged quantum metric $\det \bar{g}$ as a function of $\delta$. (c) Critical temperature $T_c$ as a function of $\delta$ for fixed system sizes ($L/a = 10^6, 10^{12}, 10^{24}$). (d) $T_c$ as a function of $L$ for $\delta = 0.01, 0.05$ and 0.1. 
\label{fig3}}
\end{figure}

As a representative example of a two-dimensional isolated flat-band system, we investigate a chiral-symmetric Lieb model~\cite{hwang2021geometric}.
The Hamiltonian for the conduction electrons in momentum space is given by
\begin{align}
H_c^{2D}(\bm{k}) = t_0 \begin{pmatrix} 0 & f_y(\bm{k}) & 0 \\ f_y^*(\bm{k}) & 0 & f_x^*(\bm{k}) \\ 0 & f_x(\bm{k}) & 0 \end{pmatrix},\end{align}
where $f_\alpha(\bm{k}) = e^{ik_\alpha a/2} + (1+\delta)e^{-ik_\alpha a/2}$ with $\delta~(>0)$ serving as a tuning parameter. 
This model yields three energy bands: a perfectly flat band at $E=0$ and two dispersive bands $E_{\pm}(\bm{k}) = \pm t_0 \sqrt{2 + 2(1+\delta)^2 + 2(1+\delta)\{\cos (k_xa) + \cos (k_ya)\}}$, as shown in Fig.~3(a).
The energy gap between the flat and dispersive bands is given by $\Delta = \sqrt{2} t_0 \delta$. The parameter $\delta$ controls both the band gap $\Delta$ and the quantum geometry of the flat band. 
We consider the regime $\Delta \gg T$, ensuring that the flat band is thermally isolated and that the ferromagnetic order of localized electrons is predominantly mediated by the flat-band states.

Crucially, the quantum metric of the flat band is highly sensitive to $\delta$. 
As shown in Fig.~\ref{fig3}(b), the determinant of the averaged quantum metric exhibits a squared-logarithmic divergence, $\det\bar g\sim[\ln(1/\delta)]^2$, as $\delta\to0$.
This geometric enhancement directly increases the RKKY-mediated spin stiffness. Consequently, the critical temperature $T_c$ rises as $\delta$ approaches zero for any fixed system size $L$, as illustrated in Fig.~\ref{fig3}(c). This trend confirms that the magnetic stability is dictated by the quantum metric tensor of the flat conduction band. 
Furthermore, Fig.~\ref{fig3}(d) displays the dependence of $T_c$ on the system size $L$ for sevaral $\delta$ values. Unlike the 1D case where $T_c \propto L^{-1/2}$, the 2D critical temperature follows a logarithmic scaling law. 
Although, $T_c$ vanishes in the thermodynamic limit ($L \to \infty$) in accordance with the Mermin-Wagner theorem, for realistic finite-size systems, the enhanced quantum metric $\bar{g}$ provides an enhanced $T_c$.

\textit{From Dispersive to Flat-Band Limits.}
The influence of quantum geometry on the RKKY interaction is not a singular phenomenon confined to the flat-band limit; rather, it represents a fundamental constituent of magnetism in all itinerant electronic systems.
To demonstrate this, we introduce a dispersion term $\epsilon(k) = -2t \cos (ka)$ into the identity component of $H_c^{1D}$, thereby endowing the system with a finite bandwidth $W = 4t$. 
By fixing the Fermi level at $0$ (half-filling) and continuously varying the ratio $W/T$, we identify two distinct physical regimes governed by the interplay between kinetic energy and quantum geometry.

In the limit where the bandwidth significantly dominates thermal energy ($W \gg T$), the system recovers classical RKKY phenomenology, where the exchange coupling follows the oscillatory decay $J_{\text{RKKY}}(r) \propto \cos(2k_F r)/r$~\cite{ruderman1954indirect,kasuya1956theory,yosida1957magnetic}. At half-filling ($k_F = \pi/2$), the strong nesting of the Fermi surface fosters dominant antiferromagnetic fluctuations at $q = \pm\pi/a$.
Notably, even in this dispersive regime, quantum geometry remains essential; it dictates the magnetic stability by substantially modulating the intensity of the susceptibility peaks, as shown in Fig.~4(a).

As the bandwidth becomes comparable to or smaller than the temperature $T$ ($W \lesssim T$), a competition emerges between the nesting-driven antiferromagnetic fluctuations and the geometry-driven ferromagnetic fluctuations. This regime suggests the possibility of a geometry-induced phase transition, where the magnetic ground state can be controlled by the geometric texture of the wavefunctions independently of the energy spectrum~\cite{oh2025magnetic,shimizu2026magnetic}.
Specifically, when $W \ll T$, the conduction electrons are thermally distributed almost uniformly across the entire Brillouin zone. In this limit, the group velocity is effectively quenched ($v_k \approx 0$), causing the dispersive contribution to vanish.
Thus, the geometric texture of the wavefunctions governs RKKY.

These results emphasize that quantum geometry is a universal factor that must be integrated into any comprehensive study of magnetism, irrespective of the energy bandwidth.

\begin{figure}[t]
\includegraphics[width=80mm]{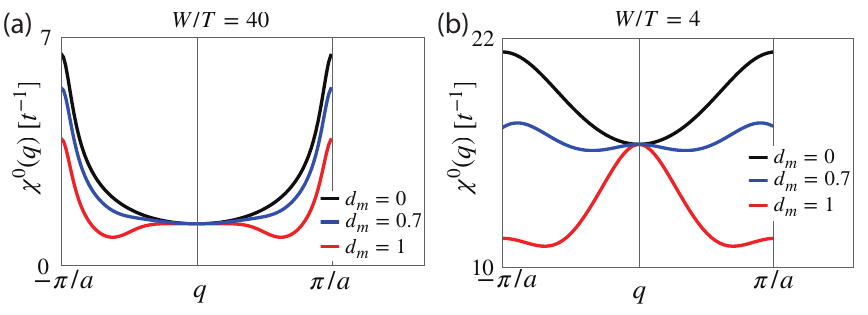} 
\caption{
\textbf{Magnetic susceptibility in a 1D dispersive band.} Static magnetic susceptibility $\chi^0(q)$ as a function of momentum $q$ for various quantum geometric parameters $d_m = \{0, 0.7, 1.0\}$. (a) $W/T=40$: Antiferromagnetic fluctuations dominate, with quantum geometry modulating the peak intensity. (b) $W/T=4$: A transition from antiferromagnetic to ferromagnetic fluctuations occurs as $d_m$ increases. All calculations are performed at half-filling ($\mu = 0$) with fixed parameters $\Delta=0.3$ and $T=0.01$.
\label{fig4}}
\end{figure}

\textit{conclusion and discussion.}
Our results demonstrate that, in the flat conduction band limit, the RKKY interaction is intrinsically determined by the quantum geometry of the Bloch states. This non-trivial geometry enables otherwise immobile electrons to mediate long-range exchange between localized moments via the finite spatial extent of their Wannier functions. Consequently, the Brillouin-zone averaged quantum metric $\bar{g}$ serves as the fundamental parameter governing both the magnetic correlation length and the spin stiffness.

The stability of this geometrically induced order is profoundly shaped by the dimensionality of the system. While the Mermin–Wagner theorem prevents long-range order in the thermodynamic limit, the geometric stiffness ($D \propto \bar{g}$) provides the necessary rigidity to suppress low-energy magnon fluctuations in finite-sized samples. Specifically, the critical temperature $T_c$ scales as $L^{-1/2} \sqrt{\bar{g}}$ in 1D and $(\ln L)^{-1/2} (\det \bar{g})^{1/4}$ in 2D, ensuring that for realistic mesoscopic dimensions, the quantum geometry can sustain a robust and observable ferromagnetic phase. 

Beyond the idealized flat-band limit, the role of quantum geometry in mediating the RKKY interaction is a universal hallmark of all itinerant electron systems.
Our findings establish a general principle for magnetic stability where the geometric form factor acts as a fundamental regulator of spin stiffness, offering a transformative perspective for engineering magnetic order across platforms ranging from flat bands to conventional dispersive magnets.
Crucially, the capacity for isospectral tuning—enhancing $T_c$ by modulating wave function geometry while preserving the energy spectrum—introduces a versatile paradigm for designing tunable, high-temperature spintronic devices. 
By revealing that magnetic rigidity can be decoupled from band dispersion, this work opens new avenues for exploring exotic magnetic phases where stability is intrinsically protected by the underlying quantum geometry.

\begin{acknowledgments}
C.~O. was supported by Japan Society for
the Promotion of Science (JSPS) KAKENHI Grant Number JP25KF0186. S. Murakami acknowledges the support by JSPS KAKENHI Grant No. JP25KF0186, NO. JP22K18687, No. JP22H00108, and No. JP24H02231. M.~S. is supported by ISHIZUE 2025 of Kyoto University. Y.~Y. is supported by JSPS KAKENHI (Grant Numbers JP22H01181, JP22H04933, JP23K17353, JP23K22452, JP24K21530, JP24H00007, JP25H01249, JP26H02016).
\end{acknowledgments}

\bibliography{ref.bib}
\end{document}